\documentclass[a4paper,aps,prl,showpacs,twocolumn,superscriptaddress]{revtex4}
\usepackage{graphicx,t1enc}

\begin{document}

\title{Applicability of the Fisher Equation to Bacterial Population Dynamics}
\author{V. M. Kenkre}
\affiliation{Consortium of the Americas for Interdisciplinary
Science and Department of Physics and Astronomy, University of New
Mexico, Albuquerque, NM 87131, U.S.A.}
\author{M. N. Kuperman}
\affiliation{Consortium of the Americas for Interdisciplinary
Science and Department of Physics and Astronomy, University of New
Mexico, Albuquerque, NM 87131, U.S.A.} \affiliation{Centro
At{\'o}mico Bariloche and Instituto Balseiro, 8400 S. C. de
Bariloche, Argentina}

\vspace{1cm}

\begin{abstract}
\ \\
\ \\
The applicability of the Fisher equation, which combines
diffusion with logistic nonlinearity, to population dynamics of
bacterial colonies is studied with the help of explicit analytic
solutions for the spatial distribution of a stationary bacterial
population under a static mask. The mask protects the bacteria
from ultraviolet light. The solution, which is in terms of
Jacobian elliptic functions, is used to provide a practical
prescription to extract Fisher equation parameters from
observations and to decide on the validity of the Fisher equation.
\end{abstract}


\maketitle

\section{ \ Introduction}

Evolution of bacterial colonies is a subject of obvious medical
importance and has been studied recently
\cite{jaco,waki,lin,nels,thesis} experimentally as well as
theoretically. Some theoretical descriptions have avoided
phenomena such as mutation and have focused on growth,
competition for resources, and diffusion. In terms of the
respective parameters $a$ (growth rate), $b$ (competition
parameter), and $D$ (diffusion coefficient), the basic equation
governing the spatio-temporal dynamics of the bacterial
population $u\left( x,t\right) $ at a position $x$ and time $t$
has been taken to be the Fisher equation \cite{fish}
\begin{equation}
\frac{\partial u\left( x,t\right) }{\partial t}=D\frac{\partial ^{2}u\left(
x,t\right) }{\partial x^{2}}+au\left( x,t\right) -bu^{2}\left( x,t\right) .
\label{originaleq}
\end{equation}
For simplicity, we consider throughout this paper only the
1-dimensional situation which is indeed appropriate to the
observations reported. Experiments have been carried out with
moving masks \cite{lin} and observations have been reported about
extinction transitions suggested earlier in theoretical
calculations \cite{nels} and in numerical simulations
\cite{thesis}. Those theoretical calculations have focused on
systems in which the growth rate $a$ varies from location to
location in a disordered manner, and have employed techniques
based on linearization of the Fisher equation. The first feature
has allowed the analysis to use concepts from Anderson localization \cite%
{anderson}, a phenomenon well-known in solid state physics of
quantum mechanical systems. The second feature has relegated the
nonlinearity character of Fisher's equation to a secondary role.
Because we suspect nonlinear features represented by $-bu^{2}$ in
(\ref{originaleq}) to be of central importance to bacterial
evolution, we have developed a theoretical approach which
generally retains the full nonlinearity of that competition term.
In the present paper, which is the first of a series built on this
approach of maintaining the nonlinearity in the equation, we
focus our
attention on the effect of a mask on the spatial distribution of the \emph{%
stationary} population of the bacteria.

Consider, as in the moving mask experiments \cite{lin}, an effectively
linear petri dish in which a mask shades bacteria from harmful ultraviolet
light which kills them in regions outside the mask but allows them to grow
in regions under the mask. Unlike in the moving mask experiments, however,
consider that the mask does not move but is left stationary. Interest is in
the $x$ -dependence of the stationary population of the bacteria. As in
previous considerations \cite{lin}, we will assume that the growth rate has
a positive constant value $a$ inside the mask, and a negative value outside
the mask.

If we take the value of $a$ outside the mask to be negative
infinite to reflect extremely harsh conditions (due to
ultraviolet light) when the bacteria are not shaded from the
light, we can take the population at the mask edges and outside
to be identically zero. We will put $\partial u\left( x,t\right)
/\partial t=0$ in (\ref{originaleq} ) to reflect stationarity,
introduce a scaled position variable $\xi =x/\sqrt{D}$ for
simplicity, and begin our analysis with the ordinary differential
equation for the stationary population $u\left( \xi \right) $:
\begin{equation}
\frac{d^{2}u\left( \xi \right) }{d\xi ^{2}}+au\left( \xi \right)
-bu^{2}\left( \xi \right) =0. \label{scaleq}
\end{equation}
Our interest is in the regions in the interior of the mask of width $2w$,
i.e., for $-w\leq x\leq w,$ the boundary conditions being $u\left( \pm w/%
\sqrt{D}\right) =0.$

The purpose of our investigation is to give a practical prescription to
decide on the applicability of the Fisher equation to bacterial evolution,
and to extract the parameters $D,\,a,\,b$ from observations if the equation
is found to be applicable.

\section{Elliptic Solutions in the Interior and Extraction of Fisher
Parameters}

The solutions of (\ref{scaleq}) can be written in terms of Jacobian elliptic
functions as follows. It is known \cite{ellip} that the square of any of cn$%
(\xi ,k)$, sn$(\xi ,k)$ or dn$(\xi ,k)$, satisfies an equation resembling (%
\ref{scaleq}). Here, we use the notation that $k$ is the elliptic
parameter \cite{defi} rather than
the elliptic modulus which is the square of $k$. Thus, $y=$ $\mbox{sn}%
^{2}(\xi ,k)$ is known to satisfy
\begin{equation}
\frac{d^{2}y}{d\xi ^{2}}+4\left( 1+k^{2}\right) y-6k^{2}y^{2}=2.  \label{y}
\end{equation}
Comparison of (\ref{y}) with (\ref{scaleq}) shows that the signs of the
linear and quadratic coefficients are the same in the two equations but (\ref%
{y}) has an extra constant term on the right hand side. This difference, as
well as the fact that the bacterial system has more independent parameters
than the single $k$ that appears in (\ref{y}), suggests that we augment $%
\mbox{sn}^{2}(x,k)$ by phase and amplitude parameters, i.e., take as the
solution of (\ref{scaleq}) within the mask
\begin{equation}
u_{i}\left( \xi \right) =\alpha \,\mbox{sn}^{2}\left( \beta \xi +\delta
,k\right) +\gamma ,  \label{solution1}
\end{equation}
and obtain the quantities $\alpha ,\beta ,\delta ,\gamma $ by
differentiating (\ref{solution1}) or by other means. The suffix $i$
represents the interior of the mask. Symmetry considerations, specifically
the requirement that the maximum of $u_{i}\left( \xi \right) $ be at $\xi
=0, $ lead to an evaluation of $\delta $ as half the period of $\mbox{sn}%
^{2} $. A shift identity allows the rewriting of (\ref{solution1}) as
\begin{equation}
u_{i}\left( \xi \right) =\alpha \,\mbox{cd}^{2}\left( \beta \xi ,k\right)
+\gamma ,  \label{soln2}
\end{equation}
the $\mbox{cd}$ function \cite{ellip} being simply the ratio $\mbox{cn}/%
\mbox{sn}.$

On differentiating (\ref{soln2}) twice w.r.t. $x$, using the relationships
among the elliptic functions, and substituting in (\ref{scaleq}), we find:
\begin{eqnarray*}
4\beta ^{2}(k^{2}+1)-a+2b\gamma  &=&0, \\
6k^{2}\beta ^{2}-b\alpha  &=&0, \\
2\alpha \beta ^{2}(1-k^{2})+\gamma (a-\gamma b) &=&0.
\end{eqnarray*}%
Solution of this algebraic system leads to the result that $\alpha $ and $%
\gamma $ are proportional to each other through a factor which is a function
only of the elliptic parameter,
\[
\gamma =\alpha \left[ \frac{-\left( k^{2}+1\right) +\sqrt{1-k^{2}+k^{4}}}{%
3k^{2}}\right] .
\]%
We also find explicit connections between the quantities $\alpha ,\beta \ $
and two of the Fisher parameters of the bacterial system $a,b,$%
\begin{equation}
\begin{array}{lll}
\alpha  & = & \left( \frac{3a}{2b}\right) k^{2}\left( 1-k^{2}+k^{4}\right)
^{-1/2} \\
\beta ^{2} & = & \left( \frac{a}{4}\right) \left( 1-k^{2}+k^{4}\right)
^{-1/2}.%
\end{array}%
  \label{alphabetagamma}
\end{equation}%
This allows us to write the stationary solution as
\begin{equation}
u_{i}\left( \xi \right) =\left( a/b\right) \left[ \,f_{\alpha }\left(
k\right) \mbox{cd}^{2}\left( \sqrt{a}f_{\beta }\left( k\right) \xi ,k\right)
+f_{\gamma }\left( k\right) \right]   \label{bestsoln}
\end{equation}%
explicitly in terms of the Fisher parameters $a,b$ and three functions of $k$
alone:
\begin{eqnarray}
f_{\alpha }\left( k\right)  &=&\left( 3/2\right) k^{2}\left( k^{\prime
2}+k^{4}\right) ^{-1/2}  \nonumber \\
f_{\beta }\left( k\right)  &=&\left( 1/2\right) \left( k^{\prime
2}+k^{4}\right) ^{-1/4}  \nonumber \\
f_{\gamma }\left( k\right)  &=&\left( 1/2\right) \left[ 1-\left(
k^{2}+1\right) \left( k^{\prime 2}+k^{4}\right) ^{-1/2}\right] .
\end{eqnarray}%
Here $k^{\prime 2}=1-k^{2}.$

Equation (\ref{bestsoln}) provides us with the means to meet the
primary goal of this investigation. The practical prescription we
seek for investigating the applicability of the Fisher equation
begins with  fitting (\ref{bestsoln}) to the observed stationary
profile. A least-squares procedure yields $a,b,k$. For
sensitivity purposes we use the nome $q=\exp \left( -\pi
K'/K\right) $ for fitting \cite{nome} rather than $k.$ The
relation
\begin{eqnarray}
\label{umax} u_{m}&=&\frac{a}{b}\left[ \,f_{\alpha }\left(
k\right) +f_{\gamma }\left( k\right) \right]\\
\nonumber
&=&\frac{a}{2b} \left( k^{2}-\,k^{\prime 2}+(k^{^{\prime
}2}+k^{4})^{1/2}\right) (k^{\prime 2}+k^{4})^{-1/2}
\end{eqnarray}%
between the maximum value of the bacterial population $u_{m}$ and
the extracted parameters provides a check on the procedure. The
determination of the diffusion constant $D$ follows the
determination of $k.$ For this we can use the boundary condition
mentioned above, that $u(\xi )$ vanishes at the edges of the mask:
$\xi =\pm w/\sqrt{D}.$ Equation ({\ref{bestsoln}) leads then to
an implicit expression which yields the diffusion constant $D:$
\begin{eqnarray}
\mbox{cn}^{2}\left( \left( a/4\right) \left( 1-k^{2}+k^{4}\right) ^{-1/2}w/%
\sqrt{D},k\right) =\, &&  \label{coli} \\
\frac{\left[ (\left( k^{2}+1\right) -\left( 1-k^{2}+k^{4}\right)
^{1/2})(1-k^{2})\right] }{k^{2}(2-k^{2}+\left(
1-k^{2}+k^{4}\right) ^{1/2})}. &&  \nonumber
\end{eqnarray}

Our prescription for the extraction of Fisher parameters $D$, $a$,
$b$ is, thus, complete provided we can assume the conditions
outside of the mask to be harsh enough to put $u$ at the edges to
vanish. This assumption can be tested from the observations. The
question of the very applicability of the Fisher equation to the
bacterial system can be addressed by the quality of the fits of
solution to the data. Fits of poor quality would necessitate a
rethinking of the quadratic nonlinearities assumed in the
equation, indeed of the entire form of the equation.

We illustrate our practical prescription in Fig. \ref{fitting}. We
have considered two hypothetical cases of the observed stationary
profile of he bacterial population. One pertains to a situation
in which the Fisher equation is applicable: (1a); the other in
which it is not: (1b). The `data' correspond, respectively,  to
stationary solutions of (\ref{originaleq}) and of the so-called
Nagumo equation \cite{kot}
\begin{equation}
\frac{\partial u}{\partial t}=D\frac{\partial ^{2}u}{\partial
x^{2}}+\left( u-C\right) \left( Au-Bu^{2}\right),   \label{nagumo}
\end{equation}
noise having been added in each case to simulate experiments.
\begin{figure}[ht]
\centering \resizebox{\columnwidth}{!} {%\rotatebox[origin=c]{-90}{
\includegraphics{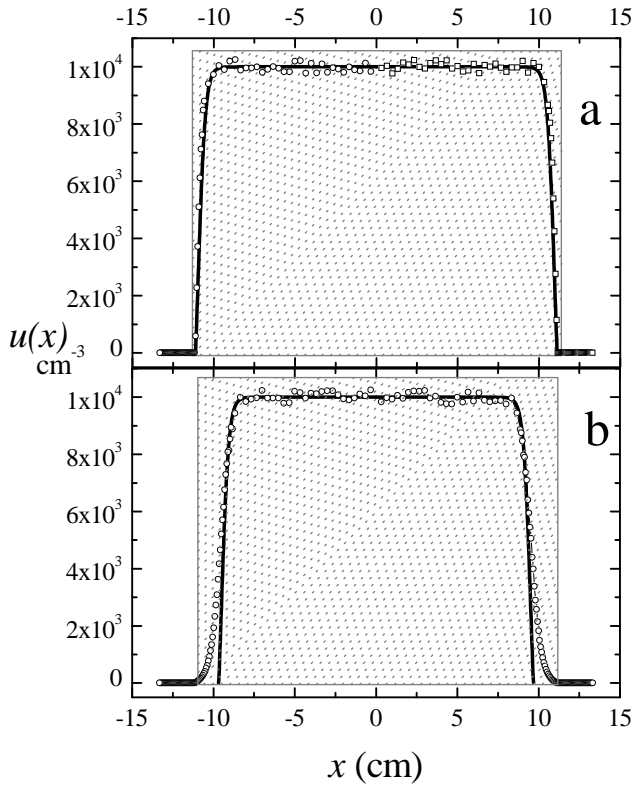}}%}
\caption{Procedure to determine applicability of Fisher equation
and/or to  extract parameters from observations. Shown is a
least-squares fitting of the analytic solution,
Eq(\ref{bestsoln}),  of the Fisher equation, to numerically
generated `data' by adding noise to theoretical predictions in two
cases. In a) the Fisher equation can be considered applicable
while in b) it cannot.} \label{fitting}
\end{figure}

The numerically generated data are plotted with circles while the
full line curve shows the best fit. We see that in Fig. 1a, the
Fisher solution matches well the data. By contrast, the fitting
procedure fails in 1b. The intrinsic non-linearities in the data
of 1b are different from those characteristic of the Fisher
equation (compare Eqs. (\ref{originaleq}) and (\ref{nagumo})).
Some of the data features in 1b, as for example the change in
concavity and the zero derivative at the borders of the mask can
not be reproduced by the analytic solution Eq (\ref{bestsoln}).
Thus, we have shown here how one would determine clearly the
applicability of the Fisher equation to a given set of
observations.

How would one proceed if, in the light of experiment, the Fisher
equation turns out to be inapplicable in this way? We suggest an
additional prescription  to obtain the form of the nonlinearity
from the stationary mask observations. The observed stationary
bacterial profile is $u_{i}\left( x\right) .$ A numerical
differentiation procedure can be made to produce
$d^{2}u_{i}\left( x\right) /dx^{2}.$ A plot of $d^{2}u_{i}\left(
x\right) /dx^{2}$ versus $u_{i}\left( x\right) ,$ the different
points corresponding to different values of $x$, would either
confirm Fisher behavior or  point to nonlinearities, such as that
in the Nagumo equation, other than that assumed in the Fisher
equation. Fig. 2 illustrates
this prescription in the context of the assumed observations in Fig. \ref%
{fitting}a and Fig. \ref{fitting}b. The `data' were numerically
differentiated in each case and the second spatial derivative was
plotted versus $u$ as shown \cite{simple}.

\begin{figure}[ht]
\centering \resizebox{\columnwidth}{!}
{\includegraphics{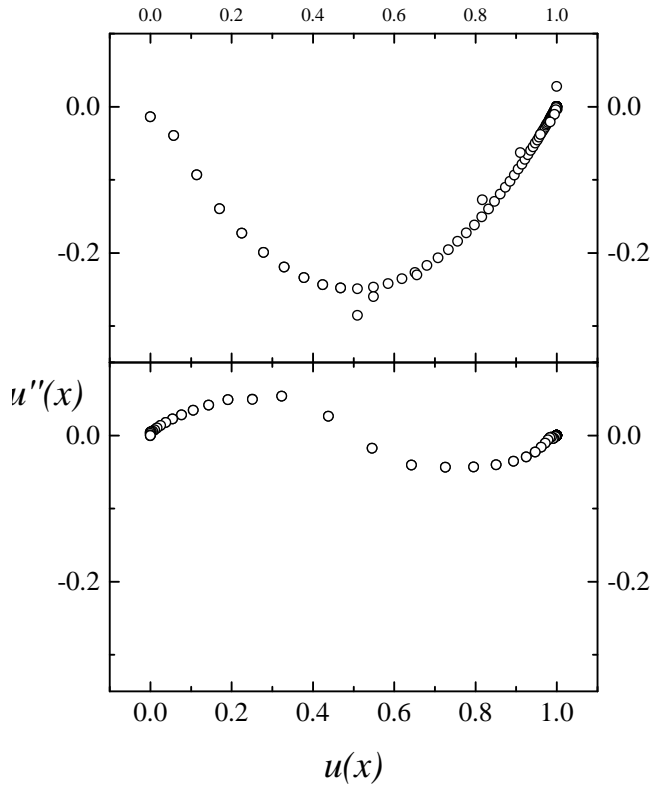}} \caption{Procedure to extract from
experiment the type of non-linearity in bacterial evolution.
Numerically obtained second derivative of $u$ is plotted against
$u$ in the two cases a and b of Fig. 1.} \label{deri2}
\end{figure}

While the quadratic nonlinearity characteristic of the Fisher
equation is compatible with Fig 2a, the curvature of the data in
Fig. 2b immediately points out the incompatibility with the Fisher
equation and suggests a Nagumo-like alternative.

\section{Dependence on Mask Size}

Obviously, good experimental practice should use for the
extraction of the Fisher parameters not a single mask but masks
of varying sizes. It is clear that the peak value of the profile,
$u_{m},$ will decrease as the mask size is decreased
(alternatively as the diffusion coefficient is increased).
However, what is the precise dependence of the stationary profile
on the size of the mask, as the size is varied? In answering this
question, one finds that the peculiarities of the elliptic
solutions produce a bifurcation behavior: there is a minimum mask
size below which bacteria cannot be supported because they
diffuse into the harsh regions where they die. We suggest that
this effect, known in the study of phytoplankton blooms
\cite{kot}, be used to validate the Fisher equation in bacterial
population as follows.

The dependence of the peak value of the stationary bacterial population on $%
k$ is in (\ref{umax}) whereas the
dependence of the mask width $2w$ $\,$on $k$ is obtained by inverting (\ref%
{coli})
\begin{equation}
\begin{array}{rcl}
w & = & \frac{\sqrt{D}}{\left( a/4\right) \left( 1-k^{2}+k^{4}\right) ^{-1/2}%
} \\
&  & \mbox{cn}^{-1}\left( \left( \frac{\left[ (\left( k^{2}+1\right) -\left(
1-k^{2}+k^{4}\right) ^{1/2})(1-k^{2})\right] }{k^{2}(2-k^{2}+\left(
1-k^{2}+k^{4}\right) ^{1/2})}\right) ^{1/2}|k\right).%
\end{array}
\label{w}
\end{equation}
The conjunction of (\ref{umax}) and (\ref{w}) yield the dependence of the
profile peak on the mask size. For a given set of Fisher parameters, a
decrease in the mask width $2w$ from large values causes a decrease in $%
k$. This decrease is monotonic. The value $k=0 $ is reached at a
finite value of the width. In this limit, the elliptic function
cd$(\beta\xi,k)$ becomes its trigonometric counterpart
cos$(\beta\xi)$, and (\ref{coli}) reduces to
\begin{equation}
\cos ^{2}\left( \frac{w}{2}\sqrt{a/D}\right) =0.5.
\end{equation}
Thus, there is a critical size $2w_{c}$ of the mask:
\begin{equation}
2w_{c}=\pi \sqrt{\frac{D}{a}}.  \label{crsize}
\end{equation}
No stationary bacterial population can be supported below such a
size. An excellent experimental check on the applicability of the
Fisher equation could be the determination of this bifurcation
behavior. On the basis of quoted  \cite{lin,thesis} values
$D\approx 10^{-5}$ cm$^{2}$/s, $a\approx 10^{-4}$/s, we obtain
the critical mask size to be of the order of half a cm, a limit
that should be observable.

\begin{figure}[ht]
\centering \resizebox{\columnwidth}{!} {\rotatebox[origin=c]{-90}{%
\includegraphics{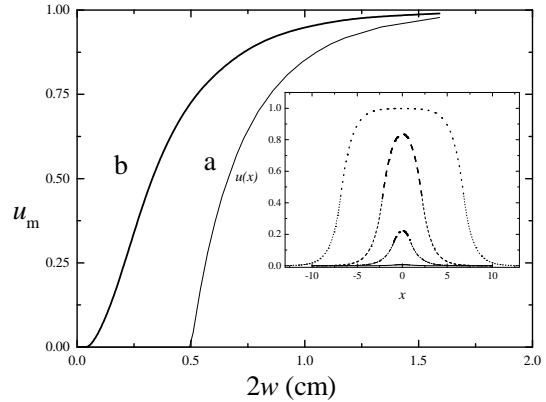}}}
\caption{Reduction of the critical size of the mask as a result
of the finiteness of $a$ outside the mask. Shown as (a) is the
dependence of the maximum of the profile, $u_m$, on the width of
the mask. For comparison we give (b), the $u_m$ dependence on
$2w$ in the Dirichlet case. For case (a), the inset shows the
actual profiles for several values of the width.} \label{diri}
\end{figure}

If we relax the condition that the environment outside the mask
is harsh enough to ensure zero population of the bacteria,
Dirichlet boundary conditions used in the previous analysis are
not appropriate. In the steady state, the bacterial concentration
just outside the borders of the mask would then be different from
zero as a result of finite diffusion. While the elliptic function
solution in Eq. (7) (but without the Dirichlet boundary condition)
is appropriate {\it inside} the mask, it turns out to be
exceedingly difficult to find a solution {\it outside} the mask.
If one starts out with the same (elliptic) form of the solution
outside but with a negative but finite value of $a,$ one gets the
requirement that $u\left( \xi\right) $ be negative. This is not
allowable since $u\left( \xi \right) $ is a bacterial density
which must remain positive. Other known solutions
\begin{equation}
u(\xi) =-\frac{\left( 3/2\right) \left( a/b\right) }{\cosh
^{2}\left( \frac{\sqrt{a}\xi}{2}\right) }
\end{equation}
are also rejected on account of their patent negativity. It is
possible, however, to obtain reasonable solutions \cite{lud} if it
is assumed that the bacterial densities outside the mask are so
small that the quadratic term proportional to $b$ may be
neglected in the Fisher equation for the analysis in the exterior
of the mask. Such an analysis leads to a smaller critical
size relative to that in (\ref{crsize}). Fig. 3 shows the dependence of $%
u_{m}$ on mask size for both the cases of (a) infinite and (b)
finite (b) (negative) $a$ outside the mask. The inset shows the
$x$ dependence of the solution for the latter case.

\section{Remarks}

Our interest in the present paper being in the determination of
the applicability of the Fisher equation to experiments currently
being conducted on the bacterial evolution in Petri dishes, we
have displayed the explicit solution (7) to the Fisher equation
(2) in the infinite-time limit when a stationary mask of given
width shades the bacteria under it from harsh conditions outside
it. Such stationary mask experiments we propose are easier and
more direct for the purposes of determination of the validity of
the Fisher equation, and for the extraction of the parameters of
the equation. It is our suggestion that parameters extracted in
this manner may be used subsequently for the analysis of moving
mask experiments \cite{lin}, with greater confidence in the
reliability of the parameter values.

We have indicated explicitly how the extraction of the Fisher
parameters may be carried out. The numerical fitting procedure in
Fig. 1a shows the parameters relevant to the hypothetical
observations to be $D=10^{-5}$cm$^2$/s , $a=10^{-4}$/s,
$b=10^{-8}$ cm$^3$/s and $w=11$ cm, while the nome $q=0.8071$
\cite{foot}. The procedure does produce parameter values  when
applied to Fig. 1b but the quality of the fits is poor. Such a
situation would signal the \emph{inapplicability} of the Fisher
equation. The `data' in Fig. 1b have been generated from the
Nagumo equation whose intrinsic nonlinearities are incompatible
with those of the Fisher equation as is visually clear from the
best fits. We have shown in Fig. 2 how general manipulations of
the observed data may be used to suggest the particular form of
nonlinearity to be used in the model. We have also concluded that
the critical size effect which arises directly from the solution
(7) is probably within observable limits for bacterial evolution,
the size we predict in light of quoted parameters being of the
order of 0.5 cm. This conclusion would necessitate modification
if the actual values of $D$ and $a$ are different from those
currently believed.

In forthcoming publications we will report our analyses of the
spatio-temporal behavior of the bacterial population of relevance
to time-dependent experiments. Our basis is the Fisher equation
\cite{KK} and several interesting alternatives such as a
formalism in which diffusion is negligible but coherent motion is
present \cite{GK}, and a formalism in which long-range
competition interactions produce an influence function and
consequently striking patterns \cite{FKK} in bacterial
populations.

\section{Acknowledgements}
This work is supported in part by the Los Alamos National
Laboratory via a grant made to the University of New Mexico
(Consortium of The Americas for Interdisciplinary Science) and by
National Science Fundation's Division of Materials Research via
grant No DMR0097204. 


\begin{thebibliography}{99}
\bibitem{nome} The quantities $K$ and $K^{\prime }$ are functions of $k$ defined as $%
K=\int_{0}^{1}\frac{dz}{\sqrt{1-z^{2}}\sqrt{1-k^{2}z^{2}}}$ and $K^{\prime
}=\int_{0}^{1}\frac{dz}{\sqrt{1-z^{2}}\sqrt{1-\left( 1-k^{2}\right) z^{2}}}.$
\bibitem{simple} Simple polynomials in $x$ were used for the fitting of the
data to assist in the numerical differentiation.
\bibitem{foot} The fitting in Fig. 1 corresponds to $w=11$ cm.
While this value appears larger than those used in current
experiments, the values of $D$ and $a$ quoted in the literature
have considerable uncertainty at the present moment.
\end{thebibliography}
\end{document}